# $Z_2$ Photonic topological insulators in the visible wavelength range for robust nanoscale photonics

*Wenjing Liu, Minsoo Hwang, Zhurun Ji, Yuhui Wang, Gaurav Modi, and Ritesh Agarwal[1]\**

Department of Materials Science and Engineering, University of Pennsylvania, Philadelphia, Pennsylvania 19104, USA

**Abstract**

Topological photonics provides an ideal platform for demonstrating novel band topology concepts, which are also promising for robust waveguiding, communication and computation applications. However, many challenges such as extremely large device footprint and functionality at short wavelengths remain to be solved which are required to make practical and useful devices that can also couple to electronic excitations in many important organic and inorganic semiconductors. In this letter, we report an experimental realization of $Z_2$ photonic topological insulators with their topological edge state energies spanning across the visible wavelength range including in the sub-500 nm regime. The photonic structures are based on deformed hexagonal lattices with preserved six-fold rotational symmetry patterned on suspended $SiN_x$ membranes. The experimentally measured energy-momentum dispersion of the topological lattices directly show topological band inversion by the swapping of the brightness of the bulk energy bands, and also the helical edge states when the measurement is taken near the topological interface. The robust topological transport of the helical edge modes in real space is demonstrated by successfully guiding circularly polarized light beams unidirectionally through sharp kinks without major signal loss. This work paves the way for small footprint photonic topological devices working in the short wavelength range that can also be utilized to couple to excitons for unconventional light-matter interactions at the nanoscale.

Arising from the demonstrated generality of band topology concepts borrowed from solid-state electronic band structures, photonic topological insulators (PTIs) exhibit topologically protected edge states that lead to unique optical transport properties such as immunity to defects and lossless propagation through sharp turns, hence providing a promising platform to build robust photonic waveguides, communication lines and circuits.[1-3] Furthermore, given the flexibility of system design and fabrication with various optical materials, PTIs serve as testbeds in demonstrating concepts that are otherwise difficult to achieve in condensed matter systems, such as Floquet topological insulators[4-7], higher order PTIs[8-14], and PT-symmetric topological structures[15-17]. PTIs have now been demonstrated in various geometries, including gratings and arrays in one dimension[18-23], photonic and plasmonic lattices and fiber arrays in two-dimensions[24-39], as well as three-dimensional PTIs[40-42], and with different mechanisms such as SSH insulators[19-23], Chern[24-27] and valley Chern insulators[28-32], and 2D and 3D $Z_2$ topological insulators[34-39]. PTIs operating at optical frequencies, especially in the *visible* wavelength range, may enable compact nanoscale topological photonic devices, as well as novel light-matter interaction phenomena. However, many of the PTI structures require either complex geometrical designs that are difficult to fabricate for shorter wavelengths, or the use of time-reversal breaking magneto-optical elements that only work up to terahertz regime, making their experimental demonstration in the visible wavelength range challenging. PTIs based on fibers, waveguides or resonator arrays can in principle work at any wavelength[4, 43], but suffer from very large device footprints and complex fabrication processes. On the other hand, $Z_2$ PTIs protected by a pseudo time-reversal symmetry do not require magneto-optical responses, which can be readily realized in photonic crystals with subwavelength structural feature sizes, hence hold promise for operation in the visible wavelength range. Recently, a PTI has been successfully demonstrated in Si pillar lattices at ~ 650 nm by Peng *et al.*[44], however, the increasing Si loss in the visible wavelength region and the fabrication complexity hinder its applications at short wavelengths. In this letter, we experimentally demonstrate PTIs with helical edge states in the visible wavelength range, tunable between 490 - 630 nm wavelengths based on $SiN_x$ photonic crystal slab. $SiN_x$ membranes are widely used in fabricating high-quality photonic devices including photonic crystals, waveguides and resonators in the visible wavelength range,[45-47] due to their low loss down until ~ 400 nm, relatively high refractive index, and mature, silicon-based fabrication techniques. Besides, the designed open cavity geometry paves the way for further coupling to numerous active materials in the visible wavelength region, to study the intriguing topological light-matter interactions.

The proposal for realizing $Z_2$ photonic topological insulators in honeycomb lattices was first put forth by Wu *et al.*,[34] and our devices are based on the specific designs proposed by Barik *et al.* based on a photonic crystal slab.[39] A similar device has been recently experimentally realized in the ~1 μm wavelength range to demonstrate certain aspects of topological photonic architectures[37, 48, 49]. An undeformed photonic honeycomb lattice, analogous to graphene, exhibits two inequivalent Dirac points located at the K and K' points of its first Brillion zone. If the unit cell is chosen as a hexagon containing six lattice sites (figure 1b, white hexagons) instead of the primitive cell (figure 1b, green rhombus), these two points will be folded to the Γ point, resulting in a doubly degenerated Dirac cone, as schematically calculated in figure 1e (obtained with a tight binding model). Deforming the lattice while keeping the $C_{6v}$ symmetry intact, i.e., decreasing (increasing) the distance between the geometrical center of each triangle and the center of the unit cell, introduces a negative (positive) mass term in the Dirac dispersion, which opens a trivial (topological) band gap. The interface between the lattices of a topological and trivial bandgap structure can therefore support helical edge states according to the bulk-edge correspondence principle. Lying within the bulk bandgap, the edge states are confined to the interface, and are protected from back scattering by a pseudo time-reversal symmetry[34], which ensures robust topological waveguiding.

The band structure and topology of the designed lattice can be understood by a simple tight binding model with the triangular holes treated as atomic sites. By considering only the nearest neighbor interactions and applying the Bloch theorem, the Hamiltonian of the hexagonal lattice containing six lattice sites can be written as[39]:

$$H = - \begin{bmatrix} 0 & t_1 & 0 & t_2\exp(i\mathbf{k}\cdot\mathbf{a_1}) & 0 & t_1 \\ t_1 & 0 & t_1 & 0 & t_2\exp(i\mathbf{k}\cdot\mathbf{a_2}) & 0 \\ 0 & t_1 & 0 & t_1 & 0 & t_2\exp(i\mathbf{k}\cdot\mathbf{a_3}) \\ t_2\exp(-i\mathbf{k}\cdot\mathbf{a_1}) & 0 & t_1 & 0 & t_1 & 0 \\ 0 & t_2\exp(-i\mathbf{k}\cdot\mathbf{a_2}) & 0 & t_1 & 0 & t_1 \\ t_1 & 0 & t_2\exp(-i\mathbf{k}\cdot\mathbf{a_3}) & 0 & t_1 & 0 \end{bmatrix}$$

(1)

where $t_1$ and $t_2$ represent the intra- and inter-cell interactions between the neighboring lattice sites (see figure 1c), respectively, $\bm{k}$ is the in-plane wave vector; $\bm{a_1}$, $\bm{a_2}$, and $\bm{a_3}$ are the three lattice vectors (see figure 1b) with $\bm{a_3} = \bm{a_2} - \bm{a_1}$. When $t_1 = t_2$, the Hamiltonian corresponds to the undeformed honeycomb lattice (analogous to graphene), giving rise to a four-fold degenerate Dirac dispersion centered at the $\Gamma$ point (figure 1e). However, a band gap opens when the lattice is deformed, with $t_1$ and $t_2$ becoming unequal (figures 1d and f). With the C$_{6v}$ symmetry preserved, the Hamiltonian can be reduced to a $4 \times 4$ matrix in the basis of $|p_\pm\rangle = |p_x\rangle \pm i|p_y\rangle$ and $|d_\pm\rangle = |d_{x^2-y^2}\rangle \pm i|d_{xy}\rangle$, corresponding to the odd (*p* orbital like) and even (*d* orbital like) parity around the $\Gamma$ point, where the + and – sign represents pseudo-spin up and down, respectively. Since the photonic energy bands involved in the designed structure are TE modes with primarily in-plane electric field, the pseudo-spin corresponds to the angular momentum of the out-of-plane **H** field, and thus the circular polarization of the in-plane **E** field. Around the $\Gamma$ point, we can expand $\bm{k}$ to the first order, i.e., $\exp(\pm i\bm{k}\cdot\bm{a_i}) \sim 1 \pm i\bm{k}\cdot\bm{a_i}$, and the pseudo-spin up and down states can be decoupled, resulting in two independent Dirac Hamiltonians in the basis of $\Psi_\pm = (|p_\pm\rangle, |d_\pm\rangle)^T$.

$$H_\pm = \frac{a}{2} t_2 (\mp k_x \sigma_x + k_y \sigma_y) + (t_2 - t_1)\sigma_z \qquad (2)$$

where $a$ is the lattice constant as defined in figure 1a, and $\sigma_x$, $\sigma_y$, and $\sigma_z$ represent the Pauli matrices. The topology of the lattice is hence determined by the relative values between $t_1$ and $t_2$, or equivalently, by the compression or expansion deformation of the lattice. In the compressed lattice, the intracell coupling $t_1$ is larger than the intercell coupling $t_2$, hence the *p* band has lower energy than the *d* band at the $\Gamma$ point (figure 1d). Upon increasing $t_2$ relative to $t_1$, the band gap closes at the point of the undeformed lattice ($t_1 = t_2$, figure 1e), and reopens for $t_2 > t_1$ with inversed *p* band and *d* band energies (figure 1f), indicating a topological phase transition. Therefore, according to the bulk edge correspondence, when the trivial and the non-trivial lattices are stitched together, the two Dirac equations, corresponding to the up ($\sigma^+$) and down ($\sigma^-$) pseudo spins, give rise to two branches of interface modes with opposite group velocities (figure 1g).

The PTI devices are composed of deformed honeycomb lattices defined by triangular holes etched on $SiN_x$ membranes (figures 1a-c and figure 2). The geometrical parameters of the $SiN_x$ photonic crystals were designed by finite difference time domain (FDTD) simulations to allow the topological band gap, and hence the edge states to sweep through the visible wavelength range. In the simulations, the unit cell, confined by Bloch boundary conditions, were excited by in-plane electric dipoles, and the energy bands were identified by a Fourier transform of the time domain response at varying in-plane k vectors $k_{||}$. Figures 2a-c present a set of band structures of the undeformed, shrunken and expanded lattices respectively calculated via FDTD simulations, with the $SiN_x$ membrane thickness of 160 nm, lattice constant, $a = 415$ nm, and the side length of the triangular hole, $d = 145$ nm. According to the simulations, the Dirac point in the undeformed lattice is located at $\lambda = 503\ nm$ for this device geometry (figure 2a). By moving the position of the triangular holes with respect to the center of the unit cell by $\pm 13\ nm$, a direct gap was opened at the Γ point with a value of 105 meV in the shrunken structure (figure 2b), and 187 meV in the expand structure (figure 2c). The spatial profiles of the $H_z$ field at the Γ point were also obtained from the simulations for the upper and lower bands (insets of figure 2b and c), with the spatial parity characteristics consistent with the qualitative prediction of the tight binding model, thereby confirming the topological phase transition. Although the upper energy band at the M point has a lower energy than at the Γ point, a complete indirect band gap is found in the expanded and shrunken lattices for 78 and 18 meV respectively, giving rise to the possibility to realize PTI devices. A larger indirect or direct bandgap at the same wavelength can be obtained with thicker structures or materials with larger dielectric constants but at a cost of smaller lateral feature sizes that increases fabrication complexities. Since the dielectric dispersion of the SiN$_x$ slab varies weakly within the 450 - 700 nm wavelength range, the energy of the bulk bands and consequently the topological edge states can be tuned nearly linearly with the lattice constant while fixing the relative ratios among all the other geometrical factors.

In our experiments, freestanding SiN$_x$ membranes with ~160 nm thickness were fabricated by KOH wet etching of patterned Si substrates with PECVD grown SiN$_x$ thin films. The photonic structures were then patterned by e-beam lithography with positive resist ZEP520A to define the etched area (triangular holes), followed by reactive ion etching of SiN$_x$. The fabricated devices were first characterized by angle-resolved reflectance measurements to identify their bulk band dispersions.[50] Figures 2d and f show the measured angle-resolved dispersion of the shrunken and expanded lattice respectively, with the center of the band gap positioned at ~ 505 nm and in-plane k vector $k_{||}$ chosen along the K' – Γ - K direction. The bulk band branches and the band gap can be clearly observed in the reflectance spectra in both lattices, which was further confirmed by the FDTD far-field simulations (figures 2e and g). The experimental data and numerically calculated results agree with each other, only with a slightly smaller band gap energy observed in the experimental data, likely due to the imperfections of the fabrication process. Importantly, owing to the spatial parity features of the PTI band structure around the Γ point, the topological band inversion can be directly observed in the spectra, indicated by an exchange of the brightness of the upper and lower energy bands at $\theta = 0°$ in the shrunken and expanded lattices respectively. In the shrunken lattice (figures 2d and e), the lower bands are *p*-orbital (dipolar) like with odd parity and non-zero dipole moments, hence appear as bright modes in the reflectance spectrum, while the upper bands are *d*-orbital (quadrupolar) like and hence dark. In the expanded lattice, on the contrary, the parity swaps between the two bands and so does the brightness of the bands in the far-field spectrum (figure 2 f and g).

After characterizing the dispersion of the bulk bands, we measured the reflectance spectrum at the interface between the shrunken and expanded lattices to directly visualize the topological edge states in the momentum space. In our structure, the trivial and topological lattices were stitched by a zig-zag interface, with the interface extending along the K' – Γ - K direction (figure 3a). When the reflectance signal was collected from the interface region with a linearly polarized excitation, two additional (linearly dispersed) modes emerge within the band gap, connecting the upper and lower bulk bands (figure 3b), indicating the existence of the topological interface states. The two interface states exhibit nearly linear dispersion within the band gap, with a crossing at the Γ point at ~500 nm. Strictly speaking, a minigap between the interface states is expected at the Γ point resulting from the $C_{6v}$ symmetry breaking at the interface, but was not resolved in our structure due to the small degree of symmetry breaking and moderate quality factor (~ 150) of the interface states. The helicity of the topological states was then confirmed by exciting the interface region with a circularly polarized light, as shown in figures 3c and d. Upon excitation with right (left) circularly polarized light, only the mode with the negative (positive) group velocity was observed, while the bulk bands show no chirality dependence, thereby proving the counter-propagating nature of the helical edge states corresponding to the pseudo-spin up and pseudo-spin down characteristics, respectively.

After confirming the existence of the edge states in the momentum space, we studied the optical propagation characteristics of the edge modes in the real space by exciting with a single wavelength laser near the topological interface with controlled polarization. Protected by the pseudo time-reversal symmetry, the helical edge states are immune to various imperfections and disorder, and can transport through sharp bends without back scattering. To demonstrate robust topological light transport, we fabricated topological interfaces with 60° and 120° turns, as shown in the bright field optical images in figure 4a, and the detailed zig-zag interface structure is shown in the SEM image in figure 4b. Upon exciting with a laser spot near the interface (Figure 4c), the topological interface lights up and successfully guided light through the designed interface kinks with minimum loss, demonstrating robust topologically protected waveguiding. To illustrate the wavelength tunability, five lattices with different lattice parameters and topological band gap centered at λ = 495, 550, 590 and 625 nm were fabricated and subsequently measured. When the wavelength of the laser was tuned within the bulk band gap of each structure, interface confined topological waveguiding was observed in all four structures (figure 4d-o). Moreover, with a linearly polarized laser (figures 4d, g, j and m), the edge modes propagate in both positive and negative $k_{\parallel}$ directions along the interface, and successfully guided light through the designed interface kinks in all the four devices with different edge state energies. Furthermore, when the laser excitation was circularly polarized (Figure 4 d-n), only the edge mode corresponding to a particular pseudo spin was excited, resulting in unidirectional propagation along the interface, consistent with the opposite group velocities of the two helical edge states observed in the momentum space dispersion (figures 3c and d). These unique properties of topological edge states give rise to possibilities of building novel photonic devices with new functionalities such as scatter-free signal transmission, spin sensitive routing, and logical operations.

To conclude, we have experimentally realized photonic topological insulators based on $SiN_x$ photonic crystal slabs that cover a large range of the visible spectrum. Helical topological interface states have been directly observed in both the real- and momentum-space, exhibiting unidirectional propagation correlated to the helicity of the excitation light. Our work demonstrates the potential to design and fabricate unconventional PTI devices with micrometer scale footprints that work in the shorter wavelength range. Moreover, the open cavity design makes it a promising platform to couple to a variety of active materials in the visible wavelength range to study novel topological light-matter interactions, and to enable actively tunable topological responses via applied optical, electrical or magnetic stimuli.[51, 52]

AUTHOR INFORMATION

**Corresponding Author**

E-mail: riteshag@seas.upenn.edu

**Notes**

The authors declare no competing financial interest.

ACKNOWLEDGEMENTS

This work was supported by the US Army Research Office under Grant No. W911NF-12-R-0012-03 and National Science Foundation under the NSF-QII-TAQS (#1936276) and NSF 2-DARE (EFMA-1542879) programs. Device fabrication work was carried out at the Singh Center for Nanotechnology, which is supported by the NSF National Nanotechnology Coordinated Infrastructure Program under grant NNCI-1542153.

# Figure captions

**Figure 1. Scheme of Z$_2$ topological photonic topological insulators.** (a)-(c) Schematic of (a) shrunken, (b) undeformed, and (c) expanded honeycomb lattices with different topology. The white hexagons outline the unit cell containing six lattice sites, while the green rhombus in (b) outlines the primitive cell in the undeformed honeycomb lattice. (d)-(f) Band structures calculated by a tight binding model corresponding to the (d) shrunken, (e) undeformed, and (f) expanded honeycomb lattices, respectively. The intra- and inter-cell interaction coefficients $t_1$ and $t_2$ used in the calculations are: (d) $t_1 = 1.05$, $t_2 = 0.95$, (e) $t_1 = t_2 = 1$, and (f) $t_1 = 0.95$, $t_2 = 1.05$. Inset of (d) and (f): the mode profile at each lattice site at the $\Gamma$ point. In (d) the lower bands exhibit odd parity and upper bands exhibit even parity, and in (d) the parity is swapped. (g) Tight binding calculation of the bulk and edge states of a finite structure composed of two regions of different topology. The edge state with positive (negative) group velocity corresponds to pseudo spin up (down), respectively.

**Figure 2. Bulk band characterization of the designed photonic crystals with different band topology.** (a)-(c) Band structures calculated by FDTD simulations, in (a) undeformed, (b) shrunken, and (c) expanded honeycomb lattices, respectively. The geometrical parameters used in the simulations are: SiN$_x$ film thickness, 160 nm, lattice constant $a = 415$ nm, side length of the triangular hole $d = 145$ nm, and the expanded and shrunken lattices are defined by moving the center of the triangular hole with respect the center of the unit cell by $\pm$ 13 nm. Insets of (b) and (c): the spatial profile of the H$_z$ field of the upper (b) and lower (c) bands at the $\Gamma$ point showing topological phase transition. (d) and (e) Experimentally measured and numerically calculated

(from FDTD) angle-resolved reflectance spectra of the shrunken lattice. (f) and (g) Experimentally measured and numerically calculated angle-resolved reflectance spectra of the expanded lattice. Scale bar of the SEM images: 400 nm.

**Figure 3. Momentum space dispersion of the topological helical interface states.** (a) Schematic of the PTI structure composed of shrunken and expanded lattice regions connected by a zig-zag domain wall. (b)-(d) Angle-resolved reflectance spectra of the topological interface state measured under (b) linearly, (c) $\sigma^-$ (clockwise circulation or right circularly) and (d) $\sigma^+$ (contour clockwise circulation or left circularly) polarized light.

**Figure 4. Real space optical waveguiding from the topological helical interface states.** (a) bright-field optical images of the fabricated PTI device composed of the shrunken and expanded lattice regions connected by a zig-zag domain wall with several 60° and 120° turns. Scale bar: 15 $\mu m$. (b) SEM image of the PTI device. Scale bar: 400 nm. (c) topological waveguiding along the interface with excitation laser wavelength of 590 nm. Scale bar: 7 $\mu m$. (d)-(o) waveguiding of the topological helical interface states at different wavelengths and with different excitation polarization. The wavelengths are: (d)-(f) 495 nm, (g)-(i) 550 nm, (j)-(l) 590 nm, (m)-(o) 625 nm, and the polarizations are: (c), (g), (j), (m) linearly polarized; (e), (h), (k), (n) $\sigma^+$ polarization; (f), (i), (l), (o) $\sigma^-$ polarization.

# Figures

**Figure 1.**

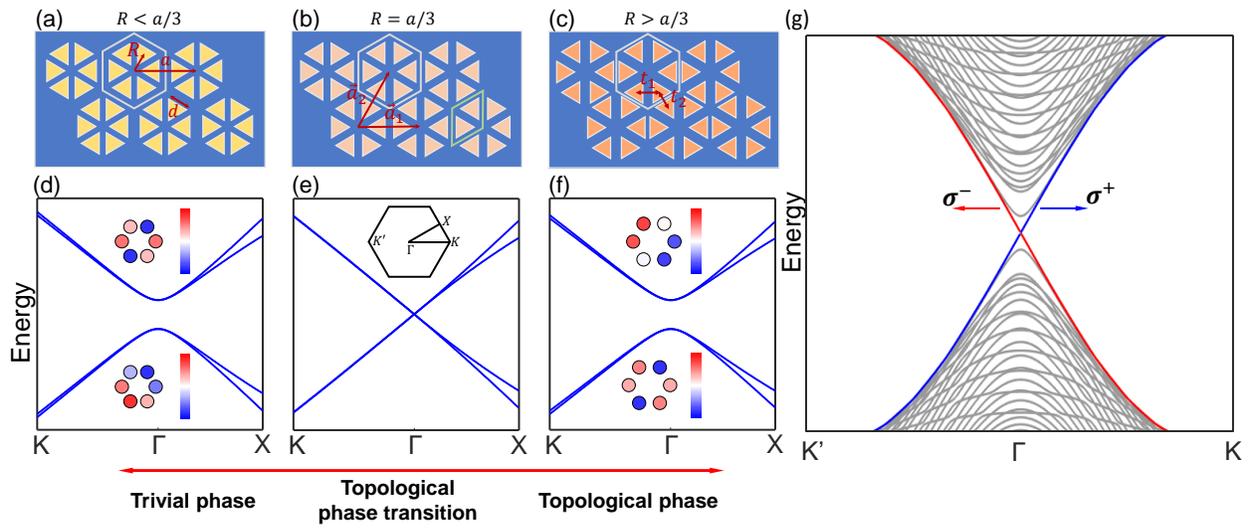

**Figure 2.**

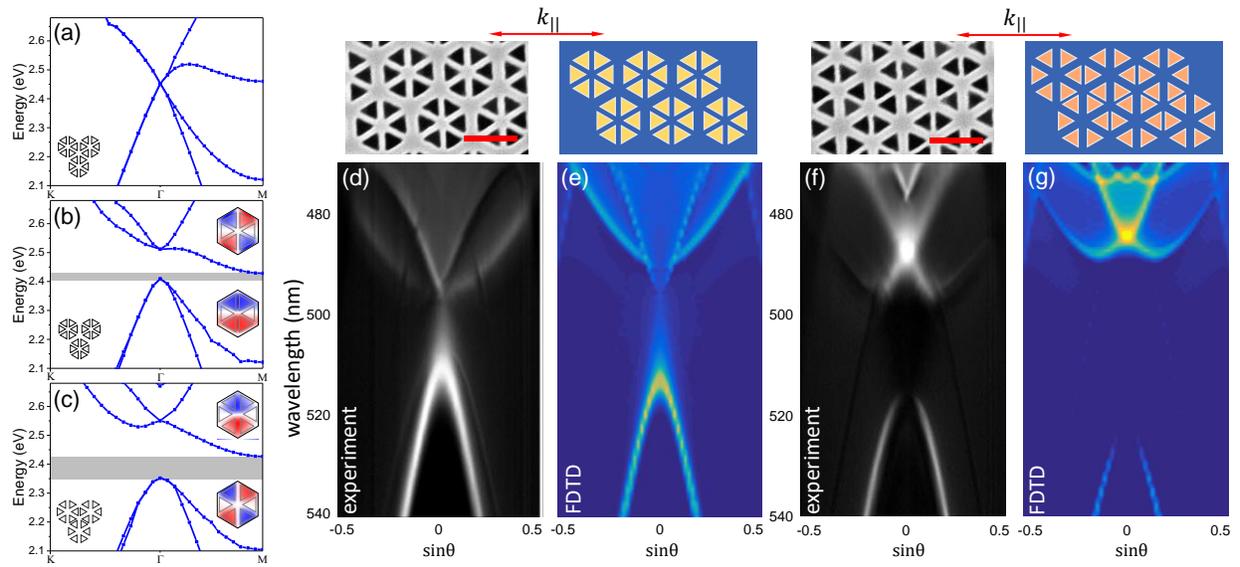

**Figure 3**.

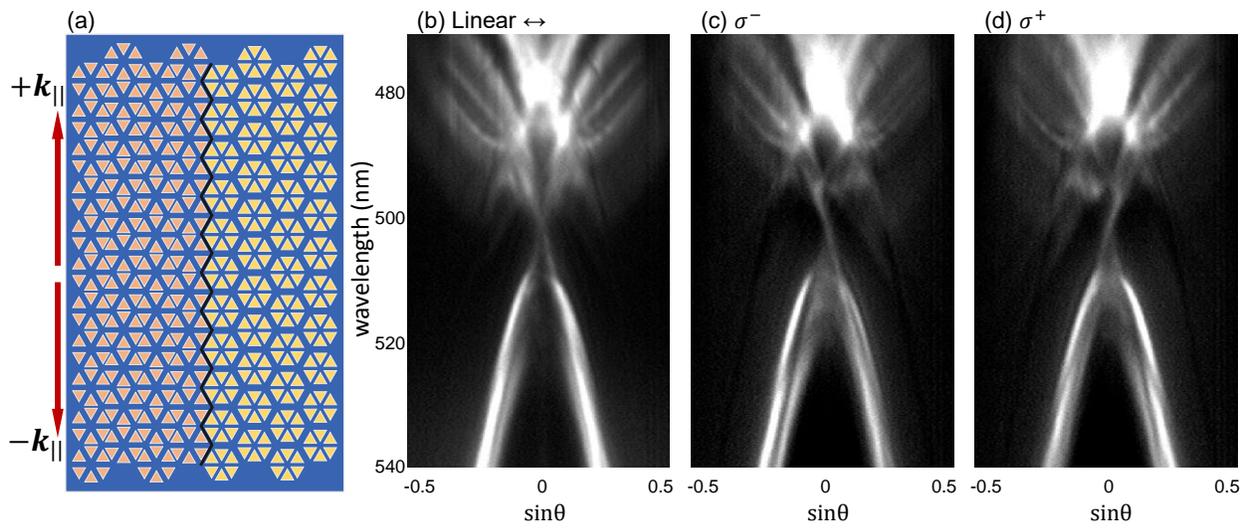

**Figure 4.**

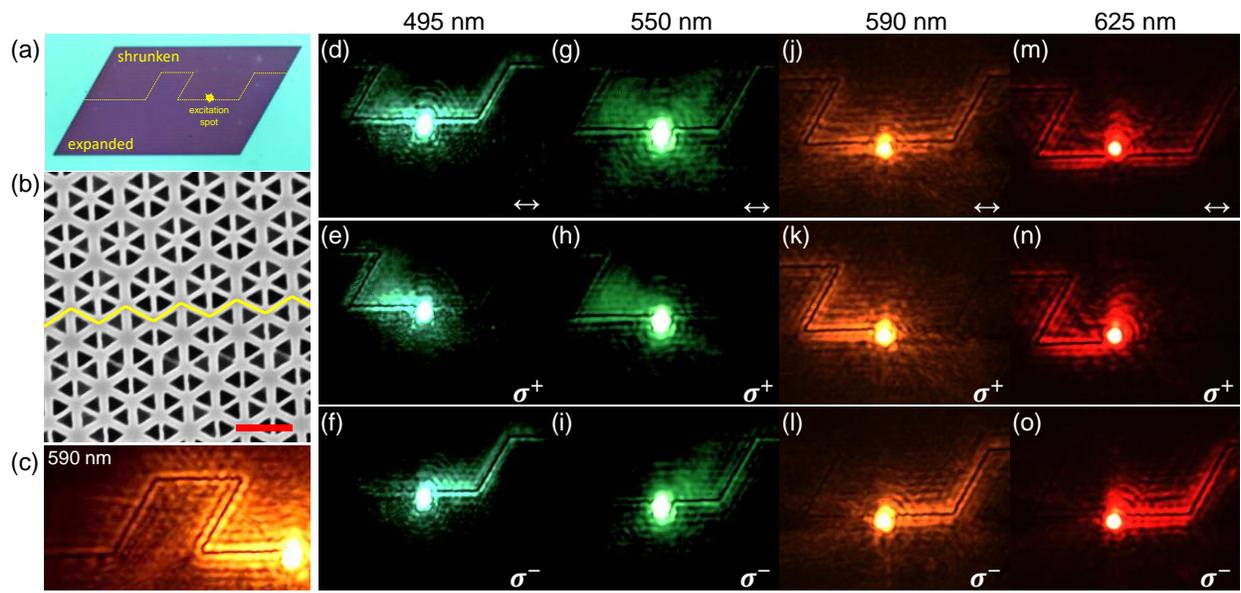

# Table of Contents Graphic

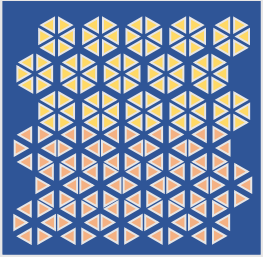 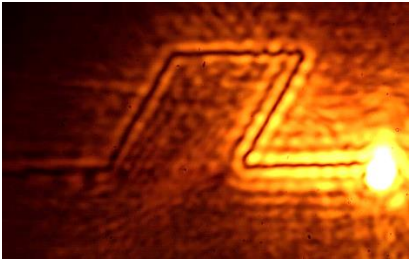